# Electrostatic Correlations and the Polyelectrolyte Self Energy

Kevin Shen and Zhen-Gang Wang[a)]
*Division of Chemistry and Chemical Engineering, California Institute of Technology
Pasadena, California, 91125, USA*

We address the effects of chain connectivity on electrostatic fluctuations in polyelectrolyte solutions using a field-theoretic, renormalized Gaussian fluctuation (RGF) theory. As in simple electrolyte solutions (Z.-G. Wang, Phys. Rev. E. **81**, 021501 (2010)), the RGF provides a unified theory for electrostatic fluctuations, accounting for both dielectric and charge correlation effects in terms of the self-energy. Unlike simple ions, the polyelectrolyte self energy depends intimately on the chain conformation, and our theory naturally provides a self-consistent determination of the response of intramolecular chain structure to polyelectrolyte and salt concentrations. The theory captures the expected scaling behavior of chain size from the dilute to semidilute regimes; by properly accounting for chain structure the theory provides improved estimates of the self energy in dilute solution and correctly predicts the eventual $N$-independence of the critical temperature and concentration of salt-free solutions of flexible polyelectrolytes. We show that the self energy can be interpreted in terms of an infinite-dilution energy $\mu_{m,0}^{el}$ and a finite concentration correlation correction $\mu^{corr}$ which tends to cancel out the former with increasing concentration.

## I. INTRODUCTION

Polyelectrolytes are widely used for many applications, ranging from energy materials[1] to solution additives (e.g. for food, cosmetics, and healthcare products).[2] Polyelectrolytes are also ubiquitous in biology, as many naturally occurring polymers – DNA/RNA, proteins, and some polysaccharides – are charged. Consequently, understanding the interplay of electrostatics with polyelectrolyte functionality is central for understanding many biological processes,[3] and guiding the design of materials such as adhesives,[4] drug-delivery microencapsulants,[5–7] and micro/nanoactuators.[8]

The long-range nature of electrostatic interactions gives rise to nontrivial correlation effects in polyelectrolyte systems such as ion-condensation[9] and complex coacervation.[10–12] A key challenge in the theoretical study of polyelectrolytes is the proper description of electrostatic correlations and their consequences on the structure and thermodynamics.

The physical origin of electrostatic correlation is the preferential interaction between opposite charges. For simple electrolyte solutions this correlation is manifested in the "ionic atmosphere" (Fig. 1) first proposed by Debye and Hückel (DH).[13] As the result of the favorable interaction of an ion with its ionic atmosphere, the free energy of the system is lowered. Theoretically, for point charges, the spatial extent of the ion atmosphere is characterized by the well-known inverse Debye screening length

$$\kappa^2 = \lambda_D^{-2} = 4\pi l_b \sum_i z_i^2 c_i, \qquad (1.1)$$

where the Bjerrum length $l_b = e^2/4\pi\varepsilon kT$ is the length scale at which two unit charges interact with energy

---

[a)] Electronic mail: zgw@caltech.edu

$kT$ and characterizes the strength of charge interactions. These charge correlations modify a host of properties such as osmotic pressure, ionic activities, and mobilities. For dilute electrolyte solutions, the electrostatic free energy density and the associated excess chemical potential are given respectively by:

$$\beta f^{el} = -\frac{\kappa^3}{12\pi}$$
$$\beta \mu_i^{el} = -z_i^2 \frac{l_b \kappa}{2}. \qquad (1.2)$$

A salient feature of the DH theory is that electrostatic correlations increase with increasing ion valency $z_i$. Thus, polyelectrolytes, being inherently multivalent, have increased correlation effects compared to simple electrolyte systems. However, the magnitude of the multivalency effect is unclear due to the spatial extent of the polyelectrolyte chains (which introduces new length scales) and the conformation degrees of freedom of the polymers (Fig. 1).

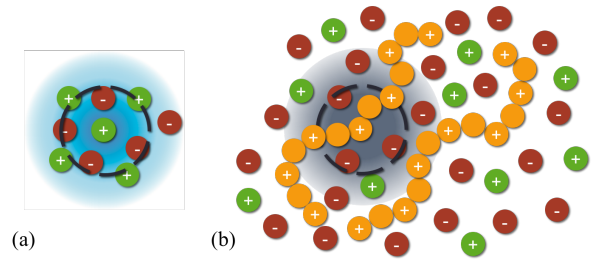

FIG. 1. (a) For simple ions, the self energy involves both a solvation contribution and, at finite concentrations, a correlation contribution due to the ionic atmosphere. (b) For polyelectrolytes, the ionic atmosphere is modified by neighboring monomers along the same and other chains, and depends on the chain structure.

The foundational (and still widely used)[14–17] theory of polyelectrolyte coacervation put forth by Voorn and Overbeek[12] completely ignores the contribution of chain connectivity to electrostatic correlations, directly borrowing the DH expressions for charge correlations and hence treating the backbone charges as disconnected free ions. The idea of approximating the backbone charge as disconnected ions has been adopted in recent theories of polyelectrolytes that replace the DH correlation using more advanced treatments of *simple* electrolytes.[18]

Another widely used theory is the thermodynamic perturbation theory (TPT), which aims to capture the effects due to chain connectivity through a leading order perturbation expansion around the simple electrolyte results.[19–24] The first-order TPT (TPT-1) forms the basis for a widely used density-functional framework for the study of inhomogeneous polyelectrolyte systems.[25–28] While producing reasonable agreements for bulk properties and inhomogeneous systems at higher densities, like the VO theory and its variants, TPT-1 offers no insight into how one can use some of the hallmark design features of polymers – backbone architecture and charge distribution on the backbone – to control electrostatic interactions in materials.

An approach that tries to capture the effects of chain connectivity from the outset is the one-loop expansion/Random Phase Approximation (RPA) based on a leading order treatment of fluctuations in the field theoretic partition function.[29–31] RPA theories require a chain structure as input, and for prescribed chain structure provide explicit expressions describing how chain-connectivity generates extra charge correlations. However, for flexible chains, theories that use a fixed, Gaussian-chain structure for all concentrations (hereafter referred to as **f**ixed **g**aussian-**RPA**) overestimate the correlation effects, particularly at dilute concentrations. As a result, for flexible chains fg-RPA predicts critical concentrations that vanish with increasing chain length,[32] in contradiction to simulation results.[33] Previous work has tried to fix this deficiency by introducing an *ad hoc* mesoscopic wavevector cut-off while still keeping the Gaussian structure factor for the long-wavelength fluctuation contributions.[34]

The RPA theories of polyelectrolytes typically only consider fluctuations in the electrostatic interactions. In order to account for excluded volume fluctuations that are important even for neutral polymer systems,[35] a variational method[36] was employed to treat the "double screening",[37] due to both excluded volume and electrostatic interactions. The theory has been used to study phase separation,[38] adapted to account for counterion condensation,[39,40] and applied to study polyelectrolyte coil-globule transitions.[41] Unlike the fg-RPA, this approach allows the polyelectrolyte conformation to self-adapt. However, in the interest of obtaining analytical expressions, the double-screening theory pre-integrates the salt degrees of freedom, resulting in the Debye-Hückel description of screened interactions between monomer units, which does not feed back on the effective interaction between the salt ions. We note that such an approximation of using a screened Coulomb interaction for macroions is a common practice in many analytical and simulation studies.[42–44]

Another method that bypasses the fixed-structure assumption is the *self-consistent* PRISM (sc-PRISM) integral equation approach[45,46] that employs a structure-dependent effective medium-induced potential; such an approach has been applied to polyelectrolytes by Yethiraj and coworkers.[43,47–50] The sc-PRISM has been quite successful for studying polyelectrolyte solution structure, but there have been limited studies on the thermodynamics,[51] perhaps due to the inconsistency between the two different routes (i.e., virial vs. compressibility) to obtaining the equation of state from the structure.[52–54] Further, PRISM theories require the use of closures, the choice of which is guided by experience and comparisons to experiment and theory.[43] Finally, to date self-consistent PRISM studies have been limited to only one polyelectrolyte species. To our knowledge the only application of PRISM to complex coacervates *ignored* the self-consistent determination of chain structure, and inconsistently borrowed thermodynamic expressions from theories of monomeric solutions.[55]

A variational theory close in spirit to the sc-PRISM was proposed by Donley, Rudnick and Liu[56] to study the concentration-dependence of polyelectrolyte chain structure. In hindsight, the theory can be understood as a sc-PRISM where the form of the effective intrachain interaction is motivated by RPA theory using heuristic arguments. Their theory yields good agreement with available computer simulation data on the end-to-end distance of the polyeletrolyte chain as a function of the polyelectrolyte and salt concentrations.

In this work, we study electrostatic fluctuations in polyelectrolyte solutions using a field-theoretic renormalized Gaussian fluctuation theory (RGF). In this theory, the key thermodynamic quantity that captures the electrostatic fluctuations is the self-energy of an effective single chain. For simple electrolytes, the self-energy is the electrostatic work required to assemble charge from an infinitely dispersed state onto an ion and is given by[57]

$$\beta\mu_{\text{chg}}^{el} = \frac{z_{\text{chg}}^2}{2}\int d\mathbf{r}'d\mathbf{r}''h_{\text{chg}}(\mathbf{r}-\mathbf{r}')G(\mathbf{r}',\mathbf{r}'')h_{\text{chg}}(\mathbf{r}''-\mathbf{r})$$
(1.3)

where $h_{\text{chg}}$ is the charge distribution on an ion, and $G(\mathbf{r},\mathbf{r}')$ is a self-consistently determined Green's function characterizing electrostatic field fluctuations, which can be thought of as an effective interaction between two test charges in an ionic environment. Eq. (1.3) is a unified expression accounting for *both* the polarization of the dielectric medium (e.g. Born solvation and image charge interactions) and correlations due to the ionic atmosphere.

For polyelectrolytes, we will see that the self energy is analogously the work required to assemble charge onto the polyelectrolyte. However, because of the conforma-tional degrees of freedom, part of the work is due to the entropic change of deforming the chain. The internal energy contribution to the self-energy resembles Eq. (1.3) and involves the single-chain structure factor, reflecting the spatial extent of the polyelectrolyte chain. The RGF theory prescribes a *self-consistent* determination of the effective intrachain structure along with the effective interaction $G(\mathbf{r}, \mathbf{r}')$ as an inherent part of the theory.

The rest of this article is organized as follows. In Section II we present a full derivation of the RGF theory for polyelectrolyte solutions. At this stage, our derivation is general for arbitrary charged macromolecules. A key part of the variational calculation is the natural emergence of a self-consistent calculation of single-chain averages and chain structure under an effective interaction $G(\mathbf{r}, \mathbf{r}')$. We then specify to a bulk system and provide expressions for the osmotic pressure and electrostatic chemical potential of polyelectrolytes, the latter being identified as the average self energy. In Section III we apply our theory to study flexible, discretely charged chains, and demonstrate how the self-consistent procedure and single-chain averages can be approximated with a variational procedure. In Section IV we present and discuss numerical results for the intrachain structure, effective interaction, polyelectrolyte self-energy, osmotic coefficient, and critical point. We compare our results with those from several existing theories, with particular attention paid to the chain length dependence in the various properties. Finally, in Section V we conclude with a summary of the key results and future outlook.

## II. GENERAL THEORY

### A. Field-Theoretic Formulation

We consider a general solution of polyelectrolytes (charged macromolecules with arbitrary internal connectivity and charge distribution) and salt ions in a solvent. We start with the microscopic density operator for species $\gamma$

$$\hat{\rho}_\gamma(\mathbf{r}) = \sum_{A=1}^{n_\gamma} \sum_{j=1}^{N_\gamma} \delta(\mathbf{r} - \mathbf{r}_{\gamma A j}), \qquad (2.1)$$

where $A$ refers to the $A$-th molecule of species $\gamma$, running up to the total number $n_\gamma$ of molecules of species $\gamma$, and $j$ refers to the $j$-th "monomer" out of a total number $N_\gamma$ of monomers in a molecule $A$ of species $\gamma$. For monomeric species, such as the small ions and solvent, $N_\gamma = 1$ and the index $j$ only takes the value of 1.

Following previous work on the self energy of simple electrolytes[57], individual charges (i.e. salt ions or charged monomers) are described by a short-ranged charge distribution $ezh(\mathbf{r}-\mathbf{r}')$, for an ion located at $\mathbf{r}'$, where we have factored out the elementary charge $e$ and the (signed) valency $z$. A convenient choice for $h$ is a Gaussian

$$h = \left(\frac{1}{2a^2}\right)^{3/2} \exp\left[-\frac{\pi(\mathbf{r}-\mathbf{r}')^2}{2a^2}\right]. \qquad (2.2)$$

This distribution gives the ion a finite radius, not of excluded volume, but of charge distribution. Doing so avoids the diverging interactions in point-charge models and captures the Born solvation energy of individual ions in a dielectric medium, as well as finite-size corrections to the ion correlation energy.[57] This feature allows our coarse-grained theory to capture the essential thermodynamic effects of finite-size ions at higher densities without having to resolve the microscopic structure.

For simple salt, the charge density of the $A$-th molecule of species $\gamma$ in unit of the elementary charge $e$ is simply $\hat{\rho}_{\gamma A}^{\mathrm{chg}} = z_\gamma h_\gamma(\mathbf{r} - \mathbf{r}_A)$. However, for polyelectrolytes we need to sum over all monomers $j$ of a particular macromolecule $A$ of species $\gamma$. In addition, to allow for arbitrary charge distribution along the polymer backbone, we introduce the signed valency $z_{\gamma,j}$ such that an uncharged monomer in the polyelectrolyte chain has $z_{\gamma,j} = 0$. The charge density due to the $A$-th molecule of the $\gamma$-th species is

$$\hat{\rho}_{\gamma A}^{\mathrm{chg}}(\mathbf{r}) = \sum_j z_{\gamma j} h_{\gamma j}(\mathbf{r} - \mathbf{r}_{\gamma A j}), \qquad (2.3)$$

where the sum runs over each monomer $j$ of object $A$ of species $\gamma$. With this definition, the charge density of each species $\gamma$ is given by

$$\hat{\rho}_\gamma^{\mathrm{chg}}(\mathbf{r}) = \sum_A \hat{\rho}_{\gamma A}^{\mathrm{chg}}. \qquad (2.4)$$

Allowing for the presence of external (fixed) charge distribution $\rho_{\mathrm{ex}}$, we then define a total charge density

$$\hat{\rho}^{\mathrm{chg}} = \rho_{\mathrm{ex}} + \sum_\gamma \hat{\rho}_\gamma^{\mathrm{chg}}. \qquad (2.5)$$

Treating the charged interactions as in a linear dielectric medium with electric permittivity $\varepsilon$ (which can be spatially dependent), the Coulomb energy of the system is written as

$$H_C = \frac{e^2}{2} \int d\mathbf{r} d\mathbf{r}' \hat{\rho}^{\mathrm{chg}}(\mathbf{r}) \mathcal{C}(\mathbf{r}, \mathbf{r}') \hat{\rho}^{\mathrm{chg}}(\mathbf{r}'), \qquad (2.6)$$

where $\mathcal{C}(\mathbf{r}, \mathbf{r}')$ is the Coulomb operator given by

$$-\nabla \cdot [\varepsilon \nabla \mathcal{C}(\mathbf{r}, \mathbf{r}')] = \delta(\mathbf{r} - \mathbf{r}'). \qquad (2.7)$$

To complete the description of the system, we add the excluded volume interactions and polyelectrolyte conformation degrees of freedom. For the excluded volume we use the incompressibility constraint in its familiar exponential representation $\delta(1 - \sum_\gamma \hat{\phi}_\gamma) = \int \mathcal{D}\eta e^{i\eta(1-\sum_\gamma \hat{\phi}_\gamma)}$, which introduces an incompressibility field $\eta$ and the volume fraction operator $\hat{\phi}_\gamma$, which is related to the density operator via $\hat{\phi}_\gamma = v_\gamma \hat{\rho}_\gamma$, where $v_\gamma$ is the monomer volume of species $\gamma$ (for simplicity, we have implicitly assumed that all monomers on the polyelectrolyte chain have the same volume). The conformation degrees of freedom of the polyelectrolyte is accounted for by an (arbitrary) chain connectivity bonded interaction hamiltonian $H_B$ which depends on the relative positions of the monomers in a polyelectrolyte chain. To focus on the electrostatic correlation, we ignore other interactions, such as the Flory-Huggins interaction.

The canonical partition function is

$$\mathcal{Q} = \prod_\gamma \frac{1}{n_\gamma! v_\gamma^{n_\gamma N_\gamma}} \prod_{A,j} \int d\mathbf{r}_{\gamma Aj} \int \mathcal{D}\eta \exp(-\beta H), \quad (2.8)$$

where the effective "Hamiltonian" is

$$\beta H = \beta H_C - i\eta \left(1 - \sum_\gamma \hat{\phi}_\gamma \right) + \beta H_B \quad (2.9)$$

and includes the incompressibility constraint. In Eq. (2.8) the species index $\gamma$ runs over all species (solvent, simple salt ions, and polyelectrolyte). We use the monomer volume $v_\gamma$ instead of the usual cube of the thermal de Broglie wavelength to avoid introducing nonessential notations; this merely produces an immaterial shift in the reference chemical potential.

We then introduce the scaled Coulomb operator

$$C = \beta e^2 \mathcal{C}$$

and the scaled permittivity

$$\epsilon = \varepsilon/(\beta e^2),$$

which has units of inverse length and is related to the Bjerrum length by

$$l_b = \frac{1}{4\pi\epsilon}.$$

For an inhomogeneous dielectric medium, $\epsilon$ will be more convenient to use than $l_b$. The scaled permittivity also leads naturally to the scaled inverse Coulomb operator

$$C^{-1}(\mathbf{r}, \mathbf{r}') = \nabla_\mathbf{r} \cdot [\epsilon(\mathbf{r})\nabla_{\mathbf{r}'}\delta(\mathbf{r} - \mathbf{r}')],$$

which is related to the Coulomb operator by

$$\int d\mathbf{r}_1 C^{-1}(\mathbf{r}, \mathbf{r}_1) C(\mathbf{r}_1, \mathbf{r}') = \delta(\mathbf{r} - \mathbf{r}')$$

To further simply notation, we henceforth use $k_B T$ as the unit of energy and $e$ as the unit of charge, so we set $\beta = 1$ and $e = 1$.

Next, we use the Hubbard-Stratanovich (HS) transformation to decouple the Coulomb interaction, which introduces the electrostatic potential $\Psi(\mathbf{r})$ (nondimensionalized by $\beta e$) and renders the canonical partition function as

$$\mathcal{Q} = \frac{1}{\Omega_C} \prod_\gamma \frac{1}{n_\gamma! v_\gamma^{n_\gamma N_\gamma}} \int \mathcal{D}\Psi \mathcal{D}\eta \exp(-Y), \quad (2.10)$$

where $\Omega_C$ is a normalization factor from the HS transformation given by

$$\Omega_C = \int D\Psi \exp\left[-\frac{1}{2}\int d\mathbf{r}d\mathbf{r}' \; \Psi(\mathbf{r}) C^{-1}(\mathbf{r},\mathbf{r}'), \Psi(\mathbf{r}')\right]$$

$$= \int D\Psi \exp\left[-\frac{1}{2}\int d\mathbf{r}\epsilon(\nabla\Psi)^2\right] = [\det C]^{1/2} \quad (2.11)$$

and the canonical "action" is

$$Y = \frac{1}{2}\int d\mathbf{r}d\mathbf{r}'\Psi(\mathbf{r}) \cdot C(\mathbf{r},\mathbf{r}')^{-1} \cdot \Psi(\mathbf{r}') \quad (2.12)$$

$$- \int d\mathbf{r}(i\eta - i\rho_{\text{ex}}\Psi) - \sum_\lambda n_\gamma \ln Q_\gamma[\eta, \Psi],$$

where $Q_\gamma$ is the single-particle/polymer partition function given shortly below.

Transforming to the grand canonical ensemble by introducing the species fugacities $\lambda_\gamma$, we obtain the grand canonical partition function

$$\Xi = \frac{1}{\Omega_C} \int \mathcal{D}\Psi \mathcal{D}\eta e^{-L[\Psi,\eta]} \quad (2.13)$$

with the grand canonical "action" $L$

$$L[\Psi,\eta] = \frac{1}{2}\int d\mathbf{r}d\mathbf{r}'\Psi(\mathbf{r}) \cdot C(\mathbf{r},\mathbf{r}')^{-1} \cdot \Psi(\mathbf{r}') \quad (2.14)$$

$$- \int d\mathbf{r}(i\eta - i\rho_{\text{ex}}\Psi) - \sum_\lambda \lambda_\gamma Q_\gamma[\eta, \Psi].$$

It is useful to notationally distinguish between the three basic types of species: solvent ($s$), simple salt ($\pm$), polymer ($p$). Correspondingly, the single-particle/polymer partition functions of the three basic types of species are:

$$Q_s = \int d\mathbf{r}_s e^{-iv_s\eta} \quad (2.15)$$

$$Q_\pm = \int d\mathbf{r}_\pm e^{-iv_\pm\eta - iz_\pm h_\pm * \Psi} \quad (2.16)$$

$$Q_p = \int \mathcal{DR} e^{-H_B - iv_p \int d\mathbf{r}\hat{\rho}_p - i\int d\mathbf{r}\Psi \hat{\rho}_{p1}^{\text{chg}}} \quad (2.17)$$

where we have introduced $\mathcal{R}$ to denote collectively the positions of all monomers in a single polyelectrolyte. For economy of notation, it is to be understood that $\hat{\rho}_{p1}^{\text{chg}}$ refers to the charge density of a *single* chain only, defined as in Eq. (2.3). We also write $h * \Psi$ to denote a convolution, or spatial averaging by the distribution $h$:

$$h * \Psi = \int d\mathbf{r}' h(\mathbf{r} - \mathbf{r}')\Psi(\mathbf{r}'). \quad (2.18)$$

Thus far Eq. (2.13) is the formally exact expression for the partition function of our system. It forms the starting point for field-based numerical simulations such as the Complex Langevin methods[58] as well as approximate analytical theories.

## B. Renormalized Gaussian Fluctuation Theory

For analytical insight, we seek to develop an approximate theory for evaluating $\Xi$, Eq. (2.13). The lowest-order saddle-point approximation would lead to a self-consistent mean-field theory,[59] in which the saddle-point condition on $\Psi$ results in a Poisson-Boltzmann (PB) level description where correlations between fixed charges $\rho_{\text{ex}}$ and mobile charges $\rho_\gamma^{\text{chg}}$ are included, but correlations between mobile charges themselves are ignored. The standard RPA theory accounts for the quadratic fluctuations around the saddle-point. In doing so, however, one is left with a *fixed* structure factor determined solely by the saddle-point condition. For uniform systems of polyelectrolytes, the chain structure factor that enters the RPA is independent of polyelectrolyte and salt concentrations. To circumvent this shortcoming, we approximate the partition function Eq. (2.13) as in our previous RGF theory, using a *non-perturbative* variational calculation.

The RGF theory follows the Gibbs-Feynman-Bogoliubov (GFB) variational approach by introducing a general Gaussian reference action $L_{\text{ref}}$. As written, Eq. (2.13) involves two fields $\eta$ and $\Psi$ to which we may apply the variational method. Allowing fluctuations in both fields will lead to the so-called double screening of both electrostatic and excluded volume.[36,37] However, in this work, we focus on the fluctuation effects due to electrostatics and thus perform the variational calculation only for the $\Psi$ field; the excluded volume interaction will be treated at the mean-field level by the saddle-point approximation for the $\eta$ field. For the $\Psi$ field, we make the following Gaussian reference action:

$$L_{\text{ref}} = \frac{1}{2} \int d\mathbf{r} d\mathbf{r}' [\Psi(\mathbf{r}) + i\psi(\mathbf{r})] G^{-1}(\mathbf{r},\mathbf{r}') [\Psi(\mathbf{r}') + i\psi(\mathbf{r}')] \tag{2.19}$$

which is parametrized by a mean electrostatic potential $-i\psi(\mathbf{r})$ and a variance, or Green's function $G(\mathbf{r},\mathbf{r}')$ which we will later show to correspond to an effective electrostatic interaction that generalizes the familiar screened-Coulomb interaction. This reference action thus accounts for the deviation $\chi = \Psi - (-i\psi) = \Psi + i\psi$ from the mean electrostatic potential.

Using $L_{\text{ref}}$ we rewrite the grand canonical partition function Eq. (2.13) as

$$\begin{aligned}\Xi &= \frac{1}{\Omega_C} \int \mathcal{D}\Psi \mathcal{D}\eta \, e^{-L_{\text{ref}}[\Psi]} e^{-(L[\Psi,\eta] - L_{\text{ref}}[\Psi])} \\ &= \frac{\Omega_G}{\Omega_C} \int \mathcal{D}\eta \left\langle e^{-(L[\Psi,\eta]-L_{\text{ref}}[\Psi])} \right\rangle_{\text{ref}} \end{aligned} \tag{2.20}$$

where $\langle \cdots \rangle_{\text{ref}}$ denotes an average over $\Psi$ with respect to the reference action $L_{\text{ref}}$, and $\Omega_G$ the corresponding partition function of $L_{\text{ref}}$, defined analogously to $\Omega_C$ in Eq. (2.11) with $G$ in place of $C$. For notational clarity, we will henceforth write $\langle \cdots \rangle_{\text{ref}}$ as $\langle \cdots \rangle$.

To implement the GFB procedure,[60] we begin with approximating the field integral over $\Psi$ with a leading order cumulant expansion

$$\Xi \approx \int \mathcal{D}\eta \, \frac{\Omega_G}{\Omega_C} e^{-\langle L - L_{\text{ref}}\rangle} \equiv \Xi_{GFB}. \tag{2.21}$$

The first cumulant in the exponent can be readily evaluated owing to the Gaussian nature of the fluctuating field, and is given by:

$$\begin{aligned}\langle L - L_{\text{ref}}\rangle &= \frac{1}{2}\int d\mathbf{r}d\mathbf{r}' \langle \Psi \cdot C^{-1} \cdot \Psi \rangle - \int d\mathbf{r}(i\eta - i\rho_{\text{ex}}\langle\Psi\rangle) - \sum_j \lambda_j \langle Q_j\rangle - \frac{1}{2}\int d\mathbf{r}d\mathbf{r}' G^{-1}(\mathbf{r},\mathbf{r}') \cdot \langle \chi(\mathbf{r})\chi(\mathbf{r}')\rangle \\ &= \frac{1}{2}\int d\mathbf{r}d\mathbf{r}' [C^{-1} - G^{-1}] \cdot G - \frac{1}{2}\int d\mathbf{r}d\mathbf{r}' \psi \cdot C^{-1} \cdot \psi - \int d\mathbf{r}(i\eta - \rho_{\text{ex}}\psi) - \sum_\gamma \lambda_\gamma \langle Q_\gamma\rangle \end{aligned} \tag{2.22}$$

where we have used $\langle\Psi\rangle = -i\psi$ and $\langle\chi(\mathbf{r})\chi(\mathbf{r}')\rangle = G(\mathbf{r},\mathbf{r}')$. The grand partition function $\Xi_{GFB}$ and *variational* grand free energy $W_v$ are found to be:

$$\Xi_{GFB} = \int \mathcal{D}\eta \, e^{-W_v[G,\psi;\eta]} \tag{2.23}$$

$$\begin{aligned}W_v[G,\psi;\eta] = &-\frac{1}{2}\ln\left(\frac{\det G}{\det C}\right) + \frac{1}{2}\int d\mathbf{r}d\mathbf{r}'[C^{-1} - G^{-1}] \cdot G \\ &- \frac{1}{2}\int d\mathbf{r}d\mathbf{r}' \psi \cdot C^{-1} \cdot \psi - \int d\mathbf{r}(i\eta - \rho_{\text{ex}}\psi) - \sum_\gamma \lambda_\gamma \langle Q_\gamma\rangle. \end{aligned} \tag{2.24}$$

In Eq. (2.24), the $\Psi$-field averaged single-particle/polymer partition functions are:

$$\langle Q_s \rangle = \int d\mathbf{r}_s \ \exp\left[-i v_s \eta\right] \tag{2.25}$$

$$\langle Q_\pm \rangle = \int d\mathbf{r}_\pm \ \exp\left[-i v_\pm \eta - z_\pm h_\pm * \psi\right] \cdot \exp\left[-\frac{1}{2} z_\pm^2 h_\pm * G * h_\pm\right] \tag{2.26}$$

$$\langle Q_p \rangle = \int \mathcal{DR} \ \exp\left[-H_B - i v_p \int d\mathbf{r} \ \eta \hat{\rho}_{p1} - \int d\mathbf{r} \ \psi \hat{\rho}_{p1}^{\text{chg}}\right] \cdot \exp\left[-\frac{1}{2}\int d\mathbf{r} d\mathbf{r}' \ \hat{\rho}_{p1}^{\text{chg}} \cdot G \cdot \hat{\rho}_{p1}^{\text{chg}}\right]. \tag{2.27}$$

For the small ions, the electrostatic fluctuations characterized by $G(\mathbf{r},\mathbf{r}')$ enter as an instantaneous self-interaction which defines the self-energy of the ion[57]

$$\begin{aligned} u_\pm(\mathbf{r}) &\equiv \frac{1}{2} z_\pm^2 h_\pm * G * h_\pm \\ &= \frac{1}{2} z_\pm^2 \int d\mathbf{r}_1 d\mathbf{r}_2 h_\pm(\mathbf{r}-\mathbf{r}_1) G(\mathbf{r}_1,\mathbf{r}_2) h_\pm(\mathbf{r}_2-\mathbf{r}). \end{aligned} \tag{2.28}$$

Similarly, the calculation of the single-chain partition function now features $G(\mathbf{r},\mathbf{r}')$ as an effective *intra*chain interaction

$$u_p^{\text{inst}}(\mathcal{R}) \equiv \frac{1}{2} \int d\mathbf{r} d\mathbf{r}' \hat{\rho}_{p1}^{\text{chg}}(\mathbf{r}) \cdot G(\mathbf{r},\mathbf{r}') \cdot \hat{\rho}_{p1}^{\text{chg}}(\mathbf{r}'). \tag{2.29}$$

In the limit where the polymer has only one monomer, $\hat{\rho}_{p1}^{\text{chg}}(\mathbf{r}) = z_p h(\mathbf{r}-\hat{\mathbf{r}})$, and the polyelectrolyte expression reduces to that of the simple electrolytes above. In the general case, however, this *instantaneous* (hence the superscript 'inst') interaction is clearly conformation dependent and *non-local*; Eq. (2.27) further suggests that there will also be chain conformation entropy contributions.

We emphasize that although our theory has the structure of independent particles and chains, the single-particle/chain partition functions involve the fluctuation-mediated effective intra-particle/chain interaction $G(\mathbf{r},\mathbf{r}')$ that is missing in self-consistent mean-field (SCMF) theories. For polymeric species, it is precisely this intrachain interaction that is able to generate chain structures that adapt to the solution conditions.

For transparency and notational simplicity, in the following we specify to a system of solvent, salt, and one polyelectrolyte species, but the expressions can be trivially extended to treat the general case with more salt and polyelectrolyte species.

To proceed, we first make the saddle-point approximation for the field $\eta$. Anticipating that the saddle-point value of $\eta$ is purely imaginary, we define a real field $\mathcal{P} = i\eta$. The saddle-point condition is

$$\frac{\delta W_v}{\delta \mathcal{P}(\mathbf{r})} = 0 \tag{2.30}$$

which yields

$$1 - v_s \rho_s - v_+ \rho_+ - v_- \rho_- - v_p \rho_p = 0. \tag{2.31}$$

The densities of the species are given by:

$$\begin{aligned} \rho_s(\mathbf{r}) &= -\frac{\lambda_p}{v_s} \frac{\delta \langle Q_s \rangle}{\delta \mathcal{P}(\mathbf{r})} = \lambda_s e^{-v_s \mathcal{P}} \\ \rho_\pm(\mathbf{r}) &= -\frac{\lambda_\pm}{v_\pm} \frac{\delta \langle Q_\pm \rangle}{\delta \mathcal{P}(\mathbf{r})} = \lambda_\pm e^{-v_\pm \mathcal{P} - z_\pm h_\pm * \psi - u_\pm} \\ \rho_p(\mathbf{r}) &= -\frac{\lambda_p}{v_p} \frac{\delta \langle Q_p \rangle}{\delta \mathcal{P}(\mathbf{r})}. \end{aligned} \tag{2.32}$$

Eq. (2.31) is just the condition of incompressibility, and $\mathcal{P}$ can be solved to yield

$$\mathcal{P} = -\frac{1}{v_s} \log \frac{1-\phi}{\lambda_s v_s} \tag{2.33}$$

where $\phi$ is the total volume fraction $\phi = \sum_{\gamma \neq s} v_\gamma \rho_\gamma$ of non-solvent species. It is customary to set $\lambda_s = 1/v_s$, which gives

$$\mathcal{P} = -\frac{1}{v_s} \log(1-\phi) \tag{2.34}$$

We note that, at the saddle-point level, our theory can be easily adapted to accommodate other models of excluded volume and hard sphere equations of state.

With the excluded volume effects taken care of, we now discuss the determination of the variational parameters $(G, \psi)$ describing the electrostatic fluctuations and interactions. The self-consistency of the GFB procedure comes from determining the values of $(G, \psi)$ such that $W_v[G, \psi; \mathcal{P}]$ is stationary at fixed pressure field $\mathcal{P}$, by a partial functional differentiation with respect to the variational parameters $G$ and $\psi$.[60,61]

Performing the variation with respect to the mean electrostatic potential

$$\frac{\delta W_v}{\delta \psi(\mathbf{r})} = 0 \tag{2.35}$$

leads to a Poisson-Boltzmann type expression

$$-\nabla \cdot \epsilon(\mathbf{r}) \nabla \psi = \rho_{\text{ex}} + \rho_+^{\text{chg}} + \rho_-^{\text{chg}} + \rho_p^{\text{chg}}, \tag{2.36}$$

where the species charge densities are given by:

$$\begin{aligned} \rho_\pm^{\text{chg}}(\mathbf{r}) &= -\lambda_\pm \frac{\delta \langle Q_\pm \rangle}{\delta \psi(\mathbf{r})} = \lambda_\pm z_\pm h_\pm * e^{-v_\pm \mathcal{P} - z_\pm h_\pm * \psi - u_\pm} \\ \rho_p^{\text{chg}}(\mathbf{r}) &= -\lambda_p \frac{\delta \langle Q_p \rangle}{\delta \psi(\mathbf{r})} \end{aligned} \tag{2.37}$$

Finally, the stationarity condition on $G$

$$\frac{\delta W_v}{\delta G(\mathbf{r}, \mathbf{r}')} = 0 \tag{2.38}$$

leads to an integro-differential equation

$$\delta(\mathbf{r} - \mathbf{r}') = \int d\mathbf{r}_1 \left[ C^{-1}(\mathbf{r}, \mathbf{r}_1) + 2I(\mathbf{r}, \mathbf{r}_1) \right] \cdot G(\mathbf{r}_1, \mathbf{r}') \tag{2.39}$$

where the ionic strength term is given by

$$\begin{aligned} 2I(\mathbf{r}, \mathbf{r}') &= \sum_\gamma \lambda_\gamma \frac{\delta \langle Q_\gamma \rangle}{\delta G(\mathbf{r}, \mathbf{r}')} \\ &= z_+^2 \int d\mathbf{r}_1 \ h_+(\mathbf{r} - \mathbf{r}_1) * \rho_+(\mathbf{r}_1) * h_+(\mathbf{r}_1 - \mathbf{r}') \\ &+ z_-^2 \int d\mathbf{r}_1 \ h_-(\mathbf{r} - \mathbf{r}_1) * \rho_-(\mathbf{r}_1) * h_-(\mathbf{r}_1 - \mathbf{r}') \\ &+ \lambda_p \langle Q_p \rangle \left\langle \hat{\rho}_{p1}^{\text{chg}} \hat{\rho}_{p1}^{\text{chg}} \right\rangle. \end{aligned} \tag{2.40}$$

In the last line above we have used the identity for the *single-chain* charge correlation

$$\left\langle \hat{\rho}_{p1}^{\text{chg}} \hat{\rho}_{p1}^{\text{chg}} \right\rangle = \frac{1}{\langle Q_p \rangle} \frac{\delta \langle Q_p \rangle}{\delta G(\mathbf{r}, \mathbf{r}')} \tag{2.41}$$

of a single chain with partition function $\langle Q_p \rangle$ to rewrite the differentiation with respect to $G$

$$\begin{aligned} \lambda_\gamma \frac{\delta \langle Q_p \rangle}{\delta G(\mathbf{r}, \mathbf{r}')} &= \lambda_p \frac{\langle Q_p \rangle}{\langle Q_p \rangle} \frac{\delta \langle Q_p \rangle}{\delta G(\mathbf{r}, \mathbf{r}')} \\ &= \lambda_p \langle Q_p \rangle \left\langle \hat{\rho}_{p1}^{\text{chg}} \hat{\rho}_{p1}^{\text{chg}} \right\rangle. \end{aligned} \tag{2.42}$$

In the case where the polymer only consists of one monomer, the polymer contribution to the ionic strength reduces to the simple electrolyte case.

Equations (2.34), (2.36), (2.39), and (2.40) constitute the central expressions of our self-consistent theory. The self-consistent determination of polymer conformation originates from the fact that the Green's function $G(\mathbf{r}, \mathbf{r}')$ Eq. (2.39) itself depends on the single-chain charge correlations $\left\langle \hat{\rho}_{p1}^{\text{chg}}(\mathbf{r}) \hat{\rho}_{p1}^{\text{chg}}(\mathbf{r}') \right\rangle$, which in turn comes from the average single-chain partition function $\langle Q_p \rangle$ determined by $G(\mathbf{r}, \mathbf{r}')$, Eq. (2.27).

Although the idea of a self-consistent determination of chain structure is not new, it is gratifying that our derivation of the RGF naturally prescribes how to perform the self-consistent calculation.

### C. Bulk Solution Thermodynamics: Self Energy and Osmotic Pressure

To demonstrate the nature of this self-consistent calculation, we now specify to a bulk solution with $\rho_{\text{ex}} = 0$.

For a bulk solution, the single-particle/polymer partition functions simplify to:

$$\begin{aligned} \langle Q_\pm \rangle &= V e^{-v_\pm \mathcal{P} - z_\pm \psi} q_\pm \\ q_\pm &= e^{-\frac{1}{2} \int dr dr' \ z_\pm h_\pm \cdot G \cdot z_\pm h_\pm} = e^{-u_\pm} \\ \langle Q_p \rangle &= V e^{-N_p v_p \mathcal{P} - z_p^{tot} \psi} q_p \\ q_p &= \frac{1}{V} \int \mathcal{DR} \ e^{-H_B - \frac{1}{2} \int dr dr' \ \hat{\rho}_{p1}^{\text{chg}} \cdot G \cdot \hat{\rho}_{p1}^{\text{chg}}} \end{aligned} \tag{2.43}$$

where $z_p^{tot}$ is the total charge carried by a chain. $q_\gamma = Q_\gamma / V$ is the single-particle/chain partition function excluding the translational degrees of freedom; for simple ions $q_\pm$ is simply the Boltzmann weight given by the simple ion self energy $u_\pm$ in Eq. (2.28).

Using Eq. (2.32), we evaluate the density and determine the fugacities to be:

$$\begin{aligned} \lambda_\pm &= \frac{n_\pm}{\langle Q_\pm \rangle} = \frac{\rho_\pm}{\exp[-u_\pm] \exp[-v_\pm \mathcal{P} - z_\pm \psi]} \\ \lambda_p &= \frac{n_p}{\langle Q_p \rangle} = \frac{\rho_p / N_p}{q_p \exp[-v_p N_p \mathcal{P} - z_p^{tot} \psi]} \end{aligned} \tag{2.44}$$

where $\rho_p = n_p N_p / V$ is the monomer density. Using the fugacity relation, the polymer contribution Eq. (2.42) to the ionic strength is simply

$$\begin{aligned} \lambda_p \langle Q_p \rangle \left\langle \hat{\rho}_{p1}^{\text{chg}} \hat{\rho}_{p1}^{\text{chg}} \right\rangle &= n_p \left\langle \hat{\rho}_{p1}^{\text{chg}} \hat{\rho}_{p1}^{\text{chg}} \right\rangle \\ &= \frac{n_p}{V} \cdot V \left\langle \hat{\rho}_{p1}^{\text{chg}} \hat{\rho}_{p1}^{\text{chg}} \right\rangle \\ &= \frac{\rho_p}{N_p} S_p^{\text{chg}}. \end{aligned} \tag{2.45}$$

Recognizing that the *single*-chain structure $\left\langle \hat{\rho}_{p1}^{\text{chg}} \hat{\rho}_{p1}^{\text{chg}} \right\rangle$ scales as the density of a single chain $N/V$, we have pre-emptively regrouped a factor of $V/V$ in anticipation that the single-chain charge structure factor defined as $S_p^{\text{chg}} \equiv V \left\langle \hat{\rho}_{p1}^{\text{chg}} \hat{\rho}_{p1}^{\text{chg}} \right\rangle$ is independent of volume.

The fugacity is related to the chemical potential by $\mu_\gamma = \ln(\lambda_\gamma v_\gamma)$, whence we can identify the per-ion and per-*chain* chemical potentials as:

$$\begin{aligned} \mu_\pm &= \ln(\rho_\pm v_\pm) + v_\pm \mathcal{P} + z_\pm \psi + u_\pm \\ \mu_p &= \ln\left(\frac{\rho_p v_p}{N_p}\right) + v_p N_p \mathcal{P} + z_p^{tot} \psi + u_p \end{aligned} \tag{2.46}$$

where

$$u_p = -\ln q_p. \tag{2.47}$$

The first three terms in both chemical potential expressions of Eq. (2.46) are the same as in a mean-field analysis of a bulk solution. The physical content of the last term $u_p$ is the free energy of a chain interacting with itself via the effective potential $G$, and within our theory $u_p$ is easily identifiable as the *per-chain*, chemical potential attributable to electrostatic fluctuations. We thus define $\mu_p^{el} \equiv u_p$ and term it the (bulk) *per-chain self-energy*.

It can be easily verified that all the polymer expressions above reduce to those of simple electrolytes in the single-monomer limit $N_p = 1$, since then $z_p^{tot} = z_p$, $H_B$ can be set to zero, $\hat{\rho}_{p1}^{chg}(\mathbf{r}) = z_p h(\mathbf{r} - \hat{\mathbf{r}})$, and by translation invariance $\int \mathcal{DR} \to \int dr \to V$. Clearly, in this limit $u_p \xrightarrow{N_p=1} u_\pm$.

In the bulk, it is also useful to define the self energy per monomer (note subscript 'm' for 'monomer') as

$$\mu_m^{el} \equiv \frac{u_p}{N_p}. \tag{2.48}$$

Further, the self-consistent set of equations (2.34), (2.36), (2.39), and (2.40) are simple in the bulk case: the constitutive equation (2.36) for $\psi$ is just the global charge neutrality constraint, while in Eq. (2.39) the structure factors and $G(\mathbf{r}, \mathbf{r}')$ become translation-invariant, allowing a simple Fourier representation. Further, because of the rotational symmetry, only the magnitude of the wavevector matters, and Eqs. (2.39) and (2.40) become:

$$1 = \epsilon k^2 \tilde{G}(k) + 2\tilde{I}(k)\tilde{G}(k) \tag{2.49}$$

$$2\tilde{I}(k) = \rho_+ \tilde{S}_+^{chg}(k) + \rho_- \tilde{S}_-^{chg}(k) + \frac{\rho_p}{N_p} \tilde{S}_p^{chg}(k) \tag{2.50}$$

which can be easily solved to obtain

$$\tilde{G}(k) = \frac{1}{\epsilon[k^2 + \tilde{\kappa}^2(k)]}, \tag{2.51}$$

where we identify $\tilde{\kappa}^2(k) = 2\tilde{I}(k)/\epsilon$ as the *wave-vector dependent* screening function, a generalization of the Debye screening constant. In our spread-charge model, even simple salt ions have some internal charge structure

$$\tilde{S}_\pm^{chg} = z_\pm^2 \tilde{h}_\pm^2(k). \tag{2.52}$$

For point charges $\tilde{h}_\pm = 1$, recovering the same ionic strength contribution as in DH theory. Therefore in the absence of polymers, in the point charge limit for simple electrolyte, $\tilde{G}(\mathbf{k})$ is precisely the DH screened Coulomb interaction.

In bulk solution, a polyelectrolyte with discrete charges has a charge structure factor that can generally be divided into a self and non-self piece

$$\tilde{S}_p^{chg}(k) = \sum_l z_{pl}^2 \tilde{h}_{pl}^2(k)$$
$$+ \sum_l \sum_{m \neq l} z_{pl} \tilde{h}_{pl}(k) z_{pm} \tilde{h}_{pm}(-k) \tilde{\omega}_{lm}(k). \tag{2.53}$$

The first sum is the $l = m$ self piece, and the second sum is over all other terms. The structure is characterized by the intramolecular correlation $\tilde{\omega}_{lm}$ between two monomers $l$ and $m$ on the same chain.[49] While $\tilde{\omega}_{lm}(k)$ has unknown analytical form, we know that $\tilde{\omega}_{lm}(k) \to 1$ as $k \to 0$, and $\tilde{\omega}_{lm}(k) \to 0$ as $k \gg 1/a$. We thus see that in the large wavelength limit the polyelectrolyte charges contribute collectively to the screening $\tilde{\kappa}^2(k)$ as a high-valency object $\sim (z^{tot})^2$ where $z^{tot} = \sum_l z_{pl}$ is the total valency. In contrast, in the small wavelength limit the charges screen as independent charges, which for historical reasons we call the Voorn-Overbeek (VO) limit (only in the sense of treating the charges as disconnected from each other – the original VO theory used DH theory with point charges, while we leave open the possibility of giving charges internal structure).

It has been previously noted that the magnitude of collective screening by polyelectrolyte charges, should be wave-vector dependent and described by the charge structure (contained in $\tilde{I}$):[30,56] at different wavelengths portions of chains screen as independent objects, and the size of these screening portions is set by the structure. These discussions correctly identified that with increasing density, screening will be increasingly controlled by higher-k structure. However, previous discussions often smear out the charges on a chain, thus treating simple ions and polymer charges on different footing and missing the approach to the the VO-limit at high wavevectors.

We now present the osmotic pressure $\Pi$. We can use $\lambda_\gamma \langle Q_\gamma \rangle / V = \rho_\gamma / N_\gamma$ to identify the ideal osmotic contribution. Then, using Fourier integrals to evaluate the determinants in $W_v$ Eq. (2.24), the osmotic pressure is:

$$\Pi = -\left[\frac{W_v - W_v^0}{V}\right] = -\frac{1}{4\pi^2} \int_0^\infty k^2 \, dk \left[\ln\left(1 + \frac{\tilde{\kappa}^2(k)}{k^2}\right) - \frac{\tilde{\kappa}^2(k)}{k^2 + \tilde{\kappa}^2(k)}\right]$$
$$- \frac{1}{v_s}\left(1 + \log(1-\phi)\right) + \frac{1-\phi}{v_s} + \rho_\pm + \frac{\rho_p}{N_p} \tag{2.54}$$

where $W_v^0$ is the grand free energy of a pure solvent system.

An important feature of the theory is the necessity of self-consistently determining the chain charge structure $\tilde{S}_p^{chg}(k)$ Eq. (2.53) and $\tilde{G}(k)$ Eq. (2.51). The Green's function $\tilde{G}(k)$ itself depends on the chain structure; the latter is in turn determined by a chain interacting with itself through $G$ in the single-chain partition function $\langle Q_p \rangle$, Eq. (2.27). The self-consistency is typically solved by iteratively approximating $G$ and the chain structure until convergence is achieved.

The last piece required to implement our theory is an evaluation of the single-chain partition function and corresponding intramolecular charge structure. The exact evaluation of single-chain partition functions is difficult even for simpler pair interactions, and in general should be done by numerical simulation.[48] In the next section we demonstrate how $\langle Q_p \rangle$ can be approximately and simply evaluated, but we point out that such an approximation is not itself inherent to the general theory.

## III. SELF-CONSISTENT CALCULATION OF FLEXIBLE CHAIN STRUCTURE

Our discussion has heretofore been general for macromolecules of arbitrary internal connectivity and charge distribution. To illustrate one way of carrying out the self-consistent calculation and facilitate comparison to previous theories, we specify to study flexible polyelectrolyte chains with Kuhn length $b$, equally spaced (discrete) charges of the same valency $z_p$, and overall charged monomer fraction $f$, such that the total polymer charge is $z_p^{tot} = Nfz_p$. Again, the discrete nature of the charges will be reflected in the charge structure factor and is important at high wavevectors.

Given this chain model, the expressions for the per-monomer chemical potential, density, and charge structure factor $\tilde{S}_\gamma^{\text{chg}}(k)$ are now:

$$\mu_m = \frac{1}{N_p} \ln \frac{\rho_p v_p}{N_p} - v_p \mathcal{P} + fz_p \psi + \frac{u_p}{N_p} \quad (3.1)$$

$$\rho_p = \lambda_p N_p e^{-N_p v_p \mathcal{P} - N_p fz_p \psi - u_p} \quad (3.2)$$

$$\tilde{S}_p^{\text{chg}}(k) = z_p^2 Nf[1 + (Nf-1)\tilde{\omega}(k)]\tilde{h}_p^2(k) \quad (3.3)$$

where we have re-expressed the sum over all monomer-monomer pair correlations $\tilde{\omega}_{lm}$ with an average per-monomer structure $\tilde{\omega}$. Following our previous discussion, one can check that when $k \to 0$, $\tilde{S}_p^{\text{chg}} \sim (Nfz_p)^2$ as for a $Nfz_p$-valent object, and when $ka \gg 1$, $\tilde{S}_p^{\text{chg}} \sim Nfz_p^2$ as for $Nf$ independent charges of valency $z_p$.

Returning to the task of calculating the single-chain partition function, we resort to a commonly used variational technique. There are many variations reported in the literature,[47,49,50,56,62–64] but they all essentially reduce to a Flory-type decomposition of the single-chain free energy into entropic $F_{\text{ent}}$ and interaction $F_{\text{int}}$ contributions

$$u_p = -\ln q_p \approx \min_\zeta \left[ F_{\text{ent}}(\zeta) + F_{\text{int}}(\zeta) \right] \quad (3.4)$$

where $\zeta$ indicates some conformational parameterization of a reference chain. In Flory's treatment of the excluded volume of a single chain, for example, $\zeta$ would be the average end-to-end distance. For our case, we take our reference chain to be a wormlike chain parameterized by an effective persistence length $\zeta \equiv l_{\text{eff}}$. Under this model, $l_{\text{eff}}$ controls the chain expansion $\alpha^2 = \langle R_{ee}^2 \rangle / R_{ee,0}^2$

$$\alpha^2 = 2\frac{l_{\text{eff}}}{b}\left[1 - \frac{l_{\text{eff}}}{Nb}\left(1 - e^{-Nb/l_{\text{eff}}}\right)\right] \quad (3.5)$$

The variational parameter $l_{\text{eff}}$ also controls the per-monomer structure $\tilde{\omega}(k; l_{\text{eff}})$ in the charge structure factor $\tilde{S}_p^{\text{chg}}$. While exact expressions of the worm-like-chain (WLC) structure factor exist in literature,[65] to facilitate calculations we use a simple analytical form:

$$\tilde{\omega}(k) = \frac{\exp[-kl_{\text{eff}}/2]}{1 + k^2 Nbl_{\text{eff}}/6} + \frac{1 - \exp[-kl_{\text{eff}}/2]}{1 + kNb/\pi} \quad (3.6)$$

which interpolates between the appropriate asymptotic limits of $\tilde{\omega}(k; l_{\text{eff}})$:[66,67]

$$\tilde{\omega}(k) \sim \begin{cases} 1, & k < \sqrt{6/Nbl_{\text{eff}}} \\ 6/k^2 Nbl_{\text{eff}}, & \sqrt{6/Nbl_{\text{eff}}} < k < 1/l_{\text{eff}} \\ \pi/kNb, & k > 1/l_{\text{eff}} \end{cases} \quad (3.7)$$

Like a previously proposed expression,[66] our expression interpolates between Gaussian-chain behavior $\tilde{\omega} \sim 6/Nbl_{\text{eff}}k^2$ at low wavevector and rodlike behavior $\tilde{\omega} \sim \pi/kNb$ at high wavevector, with a crossover set by $l_{\text{eff}}$. An important feature of the WLC chain structure captured by our expression is that the magnitude of $\tilde{\omega}$ at high wavevector is negligibly affected by $l_{\text{eff}}$, reflecting the intuition that while electrostatics can greatly deform overall chain structure, smaller scale structure is less affected,[62] consistent with blob-theory arguments[68] and simulation observations.[69] As long as we work in the regime where electrostatic blobs have only $g \sim (b/f^2 l_b)^{2/3} \sim \mathcal{O}(1)$ monomer each, the WLC structure persists down to the monomer length scale so that smaller length-scale structures do not need to be resolved.

We are thus able to write a one-parameter ($l_{\text{eff}}$) model for the single-polyelectrolyte free-energy Eq. (3.4) with entropic and interaction terms given by:

$$F_{\text{ent}} = -\frac{3}{2}N \ln\left(1 - \frac{\alpha^2}{N}\right) - 3\ln(\alpha) \quad (3.8)$$

$$F_{\text{int}} = \frac{1}{4\pi^2} \int k^2 dk \, \tilde{G}(k)\tilde{S}_p^{\text{chg}}(k; l_{\text{eff}}) \quad (3.9)$$

The first term of the entropic free energy is a finite extensibility approximation that lies between the elastic free energies obtained from integrating the worm-like-chain (WLC) and freely-jointed-chain (FJC) force-extension relationships.[70,71] The second term $-3\ln(\alpha)$ is a term that resists chain compression, first deduced by Flory and used by several subsequent authors.[72–75]

We also note that the expression for the interaction energy Eq. (3.9) is an improvement upon typical scaling estimates of the Coulomb energy. When the effective interaction $G$ is a bare Coulomb interaction, for an extended structure ($R \sim N$) the usual scaling estimate gives an energy of $N^2/R \sim N$.[68] The structure factor of an extended chain is roughly $\tilde{S}^{chg}(k) \sim N^2/(1 + kN/\pi)$, and Eq. (3.9) gives an interaction energy of $\sim N \ln N$ with the correct logarithmic correction.[76]

An interesting consequence of this decomposition of the single-chain partition function is that the electrostatic fluctuation contribution to the self energy is decomposed into two contributions: 1) entropic work of deforming the chains and 2) average interaction energy. The presence of an entropic contribution to the fluctuation-induced excess chemical potential is a special feature of flexible chains.

With the structure factor specified, we solve for self-consistency iteratively: for given Green's function $\tilde{G}$ we minimize the single-chain free energy Eq. (3.4) to approximate $l_{eff}$ and estimate the charge structure factor $\tilde{S}_p^{chg}(k)$ via Eq. (3.3) and (3.6), which we then use as input to update $\tilde{G}$ using Eq. (2.51). We stop when the rms relative error of $l_{eff}$ between iterations is below $10^{-10}$. Results of our calculations are presented in the following section.

## IV. NUMERICAL RESULTS AND DISCUSSION

For numerical calculations, we consider fully-charged chains with $f = 1$, $z_p = 1$, monovalent salt, set all ion sizes to be the same diameter $\sigma = 2a = 1$, and set the Kuhn length $b = \sigma$. We also study systems with Bjerrum length $l_b \sim 1$, ensuring that electrostatic blobs only have $g \sim (b/f^2 l_b)^{2/3} \sim 1$ Kuhn monomers each. To facilitate comparison with the restricted primitive model, the parameters $v_\gamma$ are chosen to reproduce the divergence in the excluded volume free energy at the closest packing number density of hard spheres with diameter $\sigma$. We start by examining salt-effects on the structure of isolated chains, then finite-concentration effects on chain size in salt-free polyelectrolyte solutions. Subsequently, we present the effective self-interaction $G$, demonstrate the presence of charge oscillations, and compare to screening predictions under the DH and fixed-Gaussian structure approximations. Next, we examine the polymer self-energy of *salt-free* polyelectrolyte solutions and compare our predictions to alternative theories for electrostatic correlations. Finally, we study the consequences of our theory on the osmotic coefficient and phase separation behavior.

### A. Chain Structure

The scaling behavior of linear homopolyelectrolytes is well-known for both the single-chain[68] and semidilute regimes.[77]

In the single-chain, salt-free limit, scaling theory predicts that the long range electrostatic forces elongate flexible chains into a "cigar" of electrostatic blobs with chain size scaling linearly with chain length as $R \sim N$.[68] At finite concentrations of salt, the screened Coulomb interac-tion effectively acts as an excluded volume interaction for chain segments separated by distances greater than $\kappa^{-1}$, where $\kappa$ is the inverse Debye length of the added salt. Consequently, while short chains still exhibit the "cigar" scaling, sufficiently long chains behave as self-avoiding walks,[77] with the crossover determined by the salt concentration. These expectations are borne out in Fig. 2, where we plot results for the chain size as function of chain length for several different salt concentrations $\rho_\pm$.

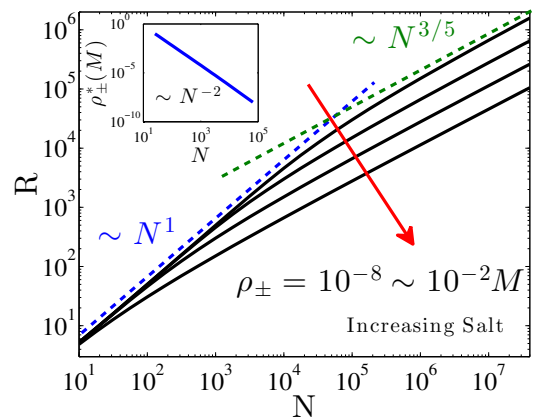

FIG. 2. End-to-end distance in the single-chain limit, with parameters $l_b = 0.7$, $f = 1$, $z_p = 1$, at different salt concentrations $\rho_\pm$. The blue and green dashed lines represent, respectively, the $R \sim N$ and $R \sim N^{3/5}$ scalings in the zero and high salt limits. The inset shows the crossover salt concentration $\rho_\pm^* \sim N^{-2}$.

For our theory, in the single-chain limit the polyelectrolyte does not contribute to the screening $\tilde{\kappa}(k)$, and the Green's function reduces to a modified screened Coulomb interaction $\tilde{G} = 4\pi l_b/(k^2 + \tilde{\kappa}(k)^2)$. Since in the single-chain limit $\tilde{G}$ is independent of polymer conformation, the self-consistent calculation only requires us to minimize the free energy of a single effective chain Eq. (3.4).

Because the salt concentrations considered are still dilute enough for finite ion-size effects to be negligible, the DH expression $\kappa^2 = 4\pi l_b z^2 (2\rho_\pm) = 8\pi l_b z^2 \rho_\pm$ is a good estimate of the screening length, and the crossover condition $\kappa R > 1$ predicts a crossover salt concentration $\rho_\pm^* \sim N^{-2}$, where we have used that in the dilute salt limit $R \sim N$. In the inset of Fig. 2 we locate the crossover by the intersection of fits to the asymptotic scaling limits, and verify this scaling expectation of the crossover concentration.

At finite concentrations of polyelectrolyte, there is a new scaling regime for the polymer size when the monomer concentration $\rho_p$ becomes sufficiently high. This concentration is usually taken to be at the physical overlap, $\rho_p^* \sim 1/N^2$, with new scaling behavior given by the semidilute prediction of ideal random walk statistics $R \sim N^{1/2}$.[77]

We plot our results for salt-free polyelectrolyte solutions in Fig. 3 at several polymer concentrations. For sufficiently high concentrations, we recover the ideal random walk scaling. Further, the crossovers happen below physical overlap, in accord with limited simulation data,[47,69] and is attributed to the fact that the Coulomb interactions are long-ranged and that chains repel each other even below physical overlap.

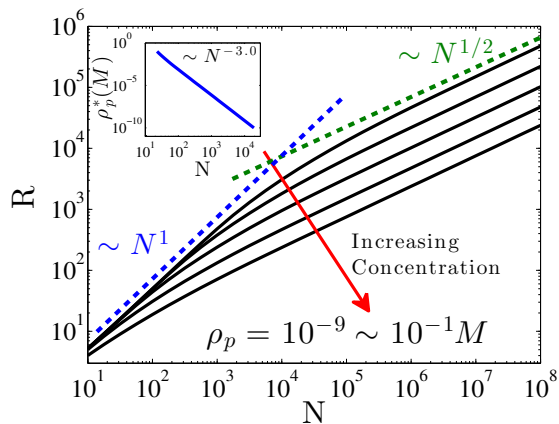

FIG. 3. End-to-end distance of a polyelectrolyte in salt-free solutions at finite monomer concentration $\rho_p$, with parameters $l_b = 0.7, f = 1, z_p = 1$. The blue and green dashed lines represent, respectively, the $R \sim N$ scaling in the dilute regime, and the $R \sim N^{1/2}$ scaling in the semidilute regime. The inset shows the crossover monomer concentration $\rho_p^* \sim N^{-3}$.

The crossover appears to be extremely gradual (more than two decades). Nevertheless, for given concentration we can approximately locate the crossover chain length by again finding the intersection of the asymptotic limits. We plot these results in the inset of Fig. 3 and find that the crossover concentration goes as $\rho_p^* \sim N^{-3}$.

To understand this apparently strong $N$-dependency, we will have to first understand the nature of screening in solutions with finite concentrations of polyelectrolyte, and we give a more detailed discussion in Section IV C. We do mention, however, that if one uses the most conservative estimate of screening where only counterions contribute to the screening length, chains are expected to interact at concentrations a factor of $1/(4\pi l_b f)^3$ below physical overlap,[77] which is several orders of magnitude for parameters studied in this paper ($4\pi l_b f \approx 10$). This is in qualitative agreement with our results that chains begin contracting far below physical overlap.

Thus, we have shown that our theory is able to correctly capture asymptotic chain size scaling behavior and reproduce qualitatively reasonable crossovers. We will show that correctly capturing these asymptotic limits is sufficient to give correct behavior for the osmotic coefficient and critical properties.

## B. Effective Interaction $G(k)$

For simple electrolytes, the field fluctuations and effective screened interaction are well-described by the Debye-Hückel screened Coulomb function. The Voorn-Overbeek approximation of neglecting chain connectivity takes Debye-Hückel as its starting point, and describes electrostatic fluctuations in polyelectrolyte solution by a screened Coulomb with screening constant $\kappa_{\text{vo}}^2 = 4\pi l_b(\rho_+ + c_p N) = 8\pi l_b \rho_p$, where the monomer number density $\rho_p$ is related to the chain number density $c_p$ by $\rho_p = c_p N$, and $\rho_+ = \rho_p$ in salt-free solution of polyelectrolyte (recall $f = 1, z_p = 1$). The opposite limit is to treat polyelectrolytes as point charges of valency $z_{tot} = N$.[69] In this case, the electrostatic fluctuations are characterized by the screened Coulomb with a renormalized screening constant $\kappa_{\text{N}}^2 = 4\pi l_b(\rho_+ + c_p N^2) = 4\pi l_b \rho_p (1 + N)$. Clearly, the effect of chain connectivity on screening should lie somewhere between these two limits. In Fig. 4, we plot the Fourier-transformed Green's function $G$ for salt-free solutions of 1) chains with adaptable structure (our theory, RGF) and for 2) chains with fixed Gaussian-chain structure (fg-RPA), and compare to the two aforementioned limits.

For sufficiently dilute systems, the Green's function $G$ for both our theory and RPA fall on the screened Coulomb line with screening constant $\kappa_{\text{N}}^2$ – information about the chain connectivity is reflected only through the total charge $z_{tot}$. We argue that this is the correct limiting law – for sufficiently dilute systems translational entropy opposes any ion condensation and the counterions can be considered a constant background charge. As long as the polymers are sufficiently far apart, they appear to each other essentially as point charges with valency $z_{tot}$, and can be treated using results from the one-component plasma (OCP) theory once one scales the charges by $z_{tot}$. In the dilute limit, the OCP is known to be governed by the DH expressions[78] – when treating polyelectrolytes as a single $z_{tot}$-valent object the OCP theory gives a screening length $\lambda$ that scales as $\sim (l_b \rho N)^{-1/2}$, which is consistent with our screening constant $\kappa_{\text{N}}^2 = 4\pi l_b \rho_p (1 + N)$ for $N \gg 1$.

At higher concentrations, the finite-extent of the polyelectrolyte begins impacting the Green's function, and leads to a peak in $G(k)$ at finite wavenumbers that depends on the concentration. This peak can be shown to lead to attractive wells in $G(\mathbf{r} - \mathbf{r}')$, which allow for positively charged chains to assume random walk statistics; the random-walk conformation would not be possible with a purely repulsive screened Coulomb interaction. This is the reason our RGF is able to reproduce the Gaussian-chain scaling in semidilute solution. On the other hand, the fg-RPA *assumes* a Gaussian-chain structure *for all concentrations*, and there is no feedback of G onto the chain structure.

The peak in $G(k)$ is also associated with decreased screening compared to DH expectations using the dilute limit $\kappa_N^2$ as the screening strength. The onset of a peak is actually also present for simple electrolyte solutions and corrects for the over-prediction of correlations within the DH approximation at higher concentrations. It can be generally shown that the peak sets in at lower concentrations for larger ion sizes: at higher concentrations, only sub-portions of spatially extended charged objects screen independently, hence decreasing the effective valency and screening strength.

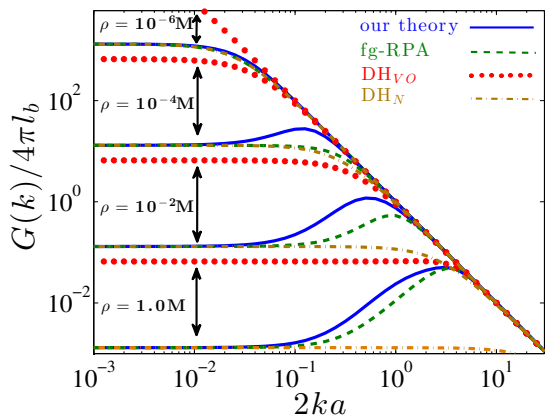

FIG. 4. Green's function $G/4\pi l_b$ characterizing electrostatic field fluctuations, at different polyelectrolyte concentrations, from our RGF theory (blue solid), fg-RPA (green dashed), DH prediction with VO screening strength $\kappa_{VO}^2 = 8\pi l_b \rho_p$ ($DH_{VO}$, red dotted), and DH prediction using the $N$-valent screening strength $\kappa_N^2 = 4\pi l_b \rho_p(1+N)$ ($DH_N$, brown dot-dashed). Results are for salt-free solutions ($l_b = 1, f = 1, z_p = 1, N = 100$) at different monomer densities $\rho_p$. Relative to the RGF, the fg-RPA over-predicts screening, has a delayed crossover, and predicts a peak at higher wavenumber (smaller wavelength).

Polyelectrolytes have their charge greatly extended across space, and correspondingly their peak sets in at a much lower concentration than simple electrolytes. The RGF, which predicts an adaptable chain size that becomes expanded relative to the ideal Gaussian chain, predicts an earlier onset of a peak in $G(k)$ and less screening (larger $G(k)$ values) than the fg-RPA theory at all concentrations. However, the peaks of the two theories do approach each other with increasing concentration, as expected of the semidilute regime.

## C. Electrostatic Self Energy and Correlations

We now examine the self energy per polyelectrolyte monomer Eq. (2.48), which depends on both the chain structure and Green's function. We first give the *total* self energy, where the zero energy of the electrostatics is taken to be the state where charges are dispersed into infinitesimal bits at infinity in vacuum. This perspective highlights the energetic consequences of connecting charges onto a chain, which is especially important in dilute solution. Further, this reference energy includes the energy of assembling charge onto each charged monomer, thus ensuring we account for both dielectric effects of solvation and interactions between charges.

To study correlation effects due to finite polymer concentration, we argue that the most natural definition involves subtracting out the infinite-dilution energy. We are then able to distinguish a dilute limit following DH-like scaling, and a crossover to less effective screening due to the overlap of polyelectrolyte chains in space.

### 1. Total Self Energy

One key feature of semidilute solutions is that the self energy should become independent of chain length for sufficiently long chain lengths. This is confirmed in Fig. 5a, where we plot the *total* self energy of salt-free solutions of polyelectrolytes. With increasing chain length, the self energies begin overlapping over greater concentration ranges, in agreement with our expectations for semidilute solutions.

Figure 5a also shows that the fg-RPA theory greatly over-estimates the self-energy in dilute solution, and rapidly grows with chain length $N$. Having shown in Section IV B that in dilute solution the Green's function $G$ becomes insensitive to chain structure, we conclude that the origin of this huge over-estimate of the self energy is the fg-RPA chain structure.

To understand the magnitude of the fg-RPA's overestimate of the self energy, we consider the infinite-dilution limit. The self energy of a simple ion is the the Born solvation energy given by $l_b/2a$, representing the work done against the dielectric background to assemble a charge into a region of size $a$. For polyelectrolytes, we expect the infinite-dilution per-monomer energy to be *higher* than the Born solvation energy of an isolated ion, due to the additional work (including chain elasticity) required to assemble multiple charges at finite separation from each other.

As confirmed in Section IV A, the chain size of an isolated flexible polyelectrolyte scales linearly with chain length $R \sim N$. Elementary calculation of the energy of a line of charges gives an energy that scales as $\sim N \ln N$; as mentioned earlier in the context of our expression for the interaction energy Eq. (3.9), this is in general true of charges arranged in a structure that scales as $R \sim N$ for large $N$. Thus for flexible polyelectrolytes we expect that at infinite dilution, the per-monomer energy associated with connectivity grows logarithmically $\sim \ln N$.

In contrast, for a fixed-Gaussian structure $R \sim N^{1/2}$ and the infinite-dilution (chain) self energy scales as $\sim N^2/R \sim N^2/N^{1/2} \sim N^{3/2}$, leading to a per-monomer self energy that grows as $\sim N^{1/2}$. This is the origin of the rapidly diverging self-energy in fg-RPA, and is attributed to the artificially compact conformation imposed by a fixed-Gaussian-chain structure. The dilute solution self energy predicted by the fg-RPA leads to an artificially high driving force for phase separation into denser states.

In contrast, our theory allows the chain conformation to relax, significantly reducing the self energy and increasing the stability of the single-phase region of a polyelectrolyte solution relative to fg-RPA theory.

For a constant dielectric background, the screening due to correlations (as a result of finite polymer concentration) reduces the amount of work required to assemble charge onto a chain, which we have defined as the self energy. The infinite-dilution self energy, then, contains information about the amount of correlation energy attributable to chain connectivity and provides an upper bound for its magnitude.

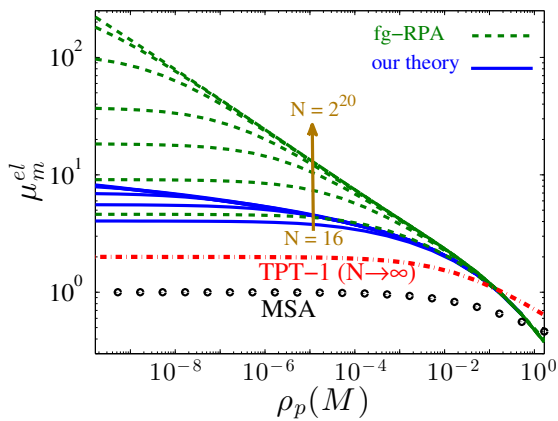

(a) Total Self Energy

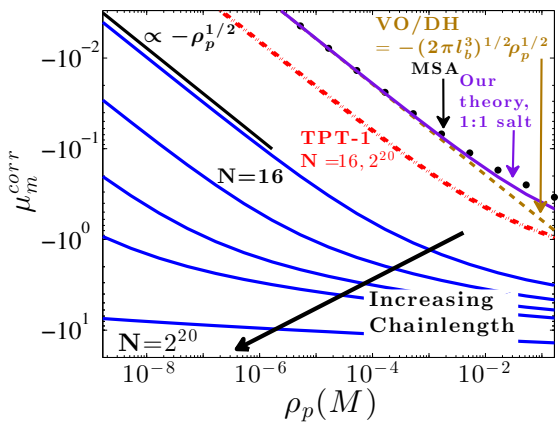

(b) Correlation Self Energy

FIG. 5. (a) Per-monomer total self energy $\mu_m^{el}$ of salt-free polyelectrolyte solutions, with parameters $l_b = 1, f = 1, z_p = 1$, comparing fg-RPA (dashed, green), RGF (solid, blue), MSA (dotted, black) and TPT-1 (dot-dashed, red) results. (b) Per-monomer correlation energy $\mu_m^{\text{corr}} = \mu_m^{el} - \mu_{m,0}^{el}$. In addition to MSA and TPT-1, we also plot the VO prediction using point-charge DH expression (dashed, brown), and our RGF theory for 1:1 salt (solid, purple). The fg-RPA correlation energy scales as $N^{1/2}$, and is omitted because it is not well-represented by the scales of the figure.

### 2. Electrostatic Correlation Energy

To isolate the correlation self energy $\mu_m^{\text{corr}}$ associated with finite concentrations of polyelectrolyte, for constant-dielectric backgrounds, it is most natural to subtract out the infinite-dilution self energy $\mu_{m,0}^{el}$

$$\mu_m^{\text{corr}} = \mu_m^{el} - \mu_{m,0}^{el} \tag{4.1}$$

In the simple-electrolyte case, this is simply taking the single-ion Born solvation energy to be the reference energy, and is the usual reference used for studying simple electrolytes. For polyelectrolytes, however, one must be careful to subtract the infinite-dilution energy of an entire chain, not just the sum of the Born solvation energy of each of the charged monomers.

In Fig. 5b we plot the correlation energy $\mu_m^{\text{corr}}$. For comparison we also plot the correlation energy results from the liquid-state integral equation Mean Spherical Approximation (MSA) theory of simple electrolytes,[24] TPT-1 chain perturbation theory,[24] the VO approximation, and our theory applied to 1:1 electrolytes. Here we focus on the behavior for our theory, and postpone comparison until Section IV C 3.

At sufficiently low concentrations, our theory predicts a per-monomer correlation energy $\mu_m^{\text{corr}}$ that scales as $\sim (N\rho_m l_b)^{1/2} = (N^2 c_p l_b)^{1/2}$, where we remind the readers that $\rho_m$ and $c_p$ are the monomer and chain number densities, respectively. Comparison with the DH point charge result $\sim (z^2 c\, l_b)^{1/2}$ indicates that in sufficiently dilute solution, the correlations follow DH-scaling, with chains screening as $N$-valent ions – the entire chain behaves as a fundamental, $N$-valent screening unit. This is in accord with the dilute limit, renormalized-DH behavior of electrostatic fluctuations described by $G(k)$.

At first sight this $N$-dependence may seem unusually strong. Examination of Fig. 5b shows that the dilute scaling quickly crosses over to a weaker concentration dependence at higher concentrations. The presence of a crossover is a generic feature of finite-sized charges, and is also present in the MSA theory for simple electrolytes and the RGF theory applied to simple electrolytes. However, the location of the crossover in simple electrolytes depends on the ion size $a$, which is much smaller than the size of a polyelectrolyte. As hinted by our examination of the chain structure and Green's function, the dilute solution DH behavior only persists while the chain size $R < \xi$, where $\xi$ is some length scale that we attempt to identify below.

In general the screening function $\hat{\kappa}$ in our RGF theory is wavelength-dependent but, as demonstrated above in the dilute solution limit, DH behavior describes the thermodynamics, with an $N$-dependent screening constant $\kappa_N^2 = 4\pi l_b \rho_p (1 + N)$, suggesting that the relevant length scale may be given by $\xi_{\text{DH}}^{-2} = \kappa_N^2$. Combined with the dilute solution scaling $R \sim N$, the condition $R/\xi_{\text{DH}} > 1$ is equivalent to $\rho_p < N^{-3} \equiv \rho_p^{*DH}$, which is the crossover scaling observed in Section IV A for the chain size. We note that this crossover scaling is in contrast to the physical overlap condition or "minimal" screening arguments that predict a crossover that scales as $\rho_p^* \sim N^{-2}$.[77] We leave the resolution of this discrepancy to future research.

Nevertheless, even if our RGF estimate of the location of crossovers is not accurate, the theory still reproduces the asymptotic limits in both dilute and semi-dilute solutions for the electrostatic correlations, both set by the scale of the infinite-dilution self energy. Thus the range of uncertainty in the intermediate concentrations is limited and therefore we expect the theory to be able to reasonably describe the thermodynamic properties in this concentration range. The same cannot be said for the fg-RPA theory, which severely overpredicts correlations in the dilute limit.

### 3. Comparison to Other Theories

We now compare our predictions for the correlation energy to other theories. Note that in Fig. 5b the fg-RPA results are not discussed because they are not well-represented on the axes used: the over-estimation of the correlation energy is too great.

As can be seen in Fig. 5b, classic VO theory approximates correlations with a solution of disconnected point charges using Debye-Hückel theory. DH theory predicts a self energy that scales linearly with the Debye screening constant $\kappa \sim \rho^{1/2}$ for all concentrations, without crossovers. Compared to our theory for polyelectrolytes, the VO theory underestimates correlations for most concentrations.

We also plot the MSA theory as an example of a liquid-state integral equation theory for the restricted primitive model of simple electrolytes, which accounts for ion size through a hard-core model. At low concentrations the integral equation theory matches the DH point charge theory; it is only at higher concentrations ($\kappa a > 1$) that there is a deviation, and correlations have a weaker concentration-dependence relative to the DH result. Recent polyelectrolyte theories that use integral equation results for the simple electrolyte[18] make the same VO approximation of treating the correlations with a solution of disconnected charges, and are expected to more or less coincide with the MSA results presented here. While for high concentrations the MSA correctly reduces correlations relative to point-charge VO, the lack of any chain-length information means that the chain-length dependent crossover is completely neglected.

Thermodynamic perturbation theory (TPT-1) is an attempt at correcting for chain correlations by perturbing about liquid-state results for simple electrolytes.[24,79] TPT-1 predicts a perturbation that grows with chain length as $\sim (N-1)/N$,[80] yielding a modest multivalency effect of chain-connectivity. However, the perturbation rapidly becomes insensitive to chain length. As shown in Fig. 5b, the TPT-1 results for chain lengths $N = 16$ and $N = 2^{20}$ are indistinguishable on the scale of the plot. Further, because the TPT-1 theory uses correlations of a simple electrolyte system, it is unable to capture the crossover behavior in the electrostatic correlations at low concentrations, an essential consequence of polymer chain connectivity. Instead, the crossover observed in TPT-1 theory is tied to the monomeric length-scale $a$.

### D. Thermodynamics and Critical Point Behavior

The theory presented in this work is applicable to the study of the thermodynamics of general polyelectrolyte solutions, which will be the subject of future work. Below we illustrate its application to the osmotic coefficients and the critical properties for a fully charged ($f = 1$) salt-free polyelectrolyte solution with monovalent counterions.

With increasing chain length $N$, we expect the osmotic coefficient to become independent of $N$ in semidilute solutions. The osmotic coefficient is defined as the ratio of the actual osmotic pressure of a solution to its ideal value (given by van't Hoff's law). For a salt-free polyelectrolyte solution with monovalent counterions, the osmotic coefficient is

$$\Phi = \frac{\Pi}{\rho_p + c_p} = \frac{\Pi}{\rho_p(1 + 1/N)} \quad (4.2)$$

In Fig. 6, we plot the RGF theory's predictions of $\Phi$ for salt-free polyelectrolyte solutions at $l_b = 1$, which can be seen to reproduce the expected convergence in the large-$N$ limit. Our result for $N = 16$ is in good quantitative agreement with reported simulation data of salt-free polyelectrolyte solutions for that chain length.[51]

Importantly, though not obvious from the figure, it can be shown that at the presented $l_b = 1$, the solution remains stable against increased chain length. Our results are in contrast to those from the fg-RPA, where at any $l_b$, increasing the chain length will eventually turn the osmotic coefficient negative and drive the system to phase separation.[32] Lastly, although the TPT-1 involves a modest chain-length correction, it far underpredicts the dependence of correlations on chain length, reflecting its behavior for the self energy.

In Fig. 7, we plot the chain-length dependence of the critical Bjerrum length ($\sim$inverse temperature) $l_b^c$ and critical monomer density $\rho_p^c$, for chain lengths up to $N = 10^4$. The predicted critical point appears insensitive to chain length for chain lengths $N \approx 30$; this insensitivity to chain length is in agreement with previous simulations and theories.[33,34]

Previous literature suggested the origin of this critical behavior as either due to counterion condensation or other strong correlations on small length scales that cannot be accounted for by weak fluctuation theories.[33,34] In contrast, our theory suggests that, in fact, accounting for chain conformational change is sufficient to explain the chain-length independence of the critical point of salt-free solutions of flexible polyelectrolytes. The conformational change is particularly important in dilute solution, where the unscreened Coulomb interactions can significantly distort the chain structure, and in doing so change the thermodynamics of phase separation.

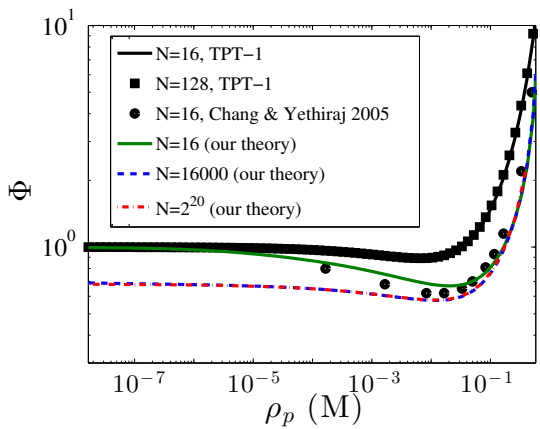

FIG. 6. Osmotic coefficient $\Phi = \Pi/\rho_p(1 + 1/N)$ of salt-free polyelectrolyte solutions as a function of the monomer concentration $\rho_p$, with parameters $l_b = 1.0, f = 1, z_p = 1$. Our RGF results for $N = 16$ (solid, green) are shown in comparison with existing simulation data (circle, black).[51] In RGF theory, as the chain length increases the osmotic coefficients approach each other at high concentration ($N = 16000$ and $N = 2^{20}$ are nearly indistinguishable). TPT-1 underpredicts the chain-length dependence, with $N = 16$ (solid, black) nearly indistinguishable from $N = 128$ (squares, black).

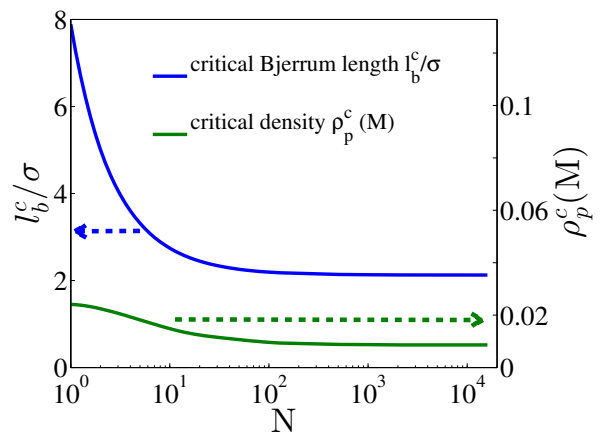

FIG. 7. Chain-length dependence of the critical Bjerrum length ($\sim 1/T^c$) and critical concentration of salt-free polyelectrolyte solutions, with parameters $f = 1, z_p = 1$.

One might notice that the predicted critical Bjerrum length for long chains $l_b^c \approx 2$ is above 1 (the critical value for Manning condensation for $f = 1$) and hence expect counterion condensation to play a role in further stabilizing the dilute phase.[33] Although counterion condensation will undoubtedly further reduce the self energy, the magnitude of such reduction is still bounded by the same infinite-dilution self energy. Because the infinite-dilution self energy per monomer has only a weak dependence on the chain length (logarithmic vs. $N^{1/2}$ from fg-RPA), the energy range in the relevant range of the concentration where counterion condensation can play a role for large $N$ in our theory is rather limited; we do not expect counterion condensation to substantially affect our conclusions. (In contrast, the magnitude of the counterion condensation contribution to the correlation energy would be much greater if the correlation energy followed the fg-RPA behavior. But even with the inclusion of counterion condensation, the fg-RPA theory does not predict the correct behavior of the critical properties without introducing additional modifications.)[34] Nevertheless, for quantitative prediction it would be important to account for counterion condensation and this is planned for future work.

## V. CONCLUSIONS

In this work, we have extended the field-theoretic renormalized Gaussian fluctuation (RGF) variational theory of simple electrolyte systems to systematically account for electrostatic fluctuations in polyelectrolytes. The key results of our theory can be summarized as follows:

1. Our theory *derives* a self-consistent procedure whereby electrostatic fluctuations characterized by $G$ Eq. (2.39) are coupled to the intrachain structure and vice versa. The theory provides a unified framework for simultaneously describing the chain structure and thermodynamics, in dilute *and* semidilute solutions.

2. Our theory correctly predicts the crossover from the $R \sim N$ scaling in the chain size to the $R \sim N^{3/5}$ scaling as a function of increasing salt in the single-chain limit. For finite concentrations of polyelectrolyte, the theory also predicts the dilute-limit scaling $R \sim N$ and the semidilute scaling $R \sim N^{1/2}$ in salt-free solutions.

3. The self-consistent procedure allows the determination of the full concentration and wavenumber dependence of the effective interaction $G$, and hence clarification of the nature of screening. We confirm the screening behavior both at long length scales, where the polyelectrolytes screen as polyvalent point charges, and smaller length scales, where the charges on the polyelectrolyte chains behave as disconnected units (the VO limit). The onset and location of a peak in $G$ is determined by the chain size, which is more accurately described by an adaptive chain structure.

4. Our theory features prominently the role of the polyelectrolyte self-energy $u_p$, Eq. (2.47), which is the free energy of an independent chain interacting with itself through $G$, and is the work required to assemble charge onto a chain. The infinite-dilution self-energy bounds the magnitude of connectivity contributions to the correlation energy, the latter tending to cancel out the former with increasing concentration.

5. We clarify that the correlation energy $\mu_m^{\text{corr}}$ is the difference of the self energy from its infinite-dilution value; $\mu_m^{\text{corr}}$ characterizes finite-concentration effects and reduces the self energy. In sufficiently dilute solutions, $\mu_m^{\text{corr}}$ follows a *universal* renormalized DH scaling $\mu_m^{\text{corr}} \sim -(\rho_p N)^{1/2}$, independent of chain structure. Above some chain-size dependent crossover, chain connectivity results in a weaker concentration dependence.

6. By predicting the correct infinite-dilution chain structure and self energy, for salt-free solutions our theory captures the $N$-insensitivity both of the osmotic coefficient in semidilute solution and the critical properties in the large-$N$ limit.

We note that our physical picture of the self-energy corroborates the self-energy explanation used by some authors for "strong correlation" complexation.[81–83] These works treated polyelectrolyte complexation in the zero temperature limit where Coulomb interactions dominate. They identified the driving force for polyelectrolytes to aggregate into denser states as driven by a loss of an infinite-dilution self energy (which was estimated using the Coulomb energy of line charges $\sim N \log N$ per chain) upon entering a dense, neutralized state. Our theory works at finite temperature, and for a given concentration is able to *quantify* how much of the infinite-dilution self energy is lost. Being a weak coupling theory, our theory will require further modification to include structures due to strong correlation effects, such as counterion condensation and ion-pairing.

The response of chain conformation to changing density is a key feature in our theory that is not present in theories of polyelectrolytes assuming fixed chain structures, such as fg-RPA[32,34] or Ultra-Soft-Restricted-Primitive Model.[84,85] Such theories predict, for Gaussian chains, spuriously strong $N$-dependencies of correlation energies, and this behavior is due the failure of the assumed Gaussian structure in dilute concentrations. The fixed-Gaussian structure assumption artificially confines flexible chains to a radius that is too small by a factor of $\sqrt{N}$, thus raising the infinite-dilution self energy by the same factor. Thus, the fg-RPA theory predicts a higher infinite-dilution energy (which is a positive quantity), and there is correspondingly more electrostatic energy for the correlations (which contribute a negative energy) to reduce. By a self-consistent determination of the single-chain structure, our theory avoids the artificially high energies appearing in the fg-RPA theory.

For context, our theory's requirement of self-consistency between an effective interaction $G$ and a single-chain structure is similar in spirit to sc-PRISM proposals[47] and the procedure proposed by Donley et al.[56] Indeed, if one uses the PRISM equations with the so-called "RPA" closure,[47,86] the smeared charge distributions $h$ to regularize the electrostatic interactions, and a commonly employed estimate of the medium-induced potential,[45–47] one will recover the same total effective interaction as our Green's function $G$.

However, in contrast to sc-PRISM our theory comes furnished with expressions for thermodynamics, Eq. (2.24), (2.46). We also emphasize that, in addition to the *intra*chain structure presented in this work, our theory should also be able to predict an *inter*chain structure that goes *beyond* that predicted by the aforementioned procedure of using sc-PRISM with the "RPA" closure. This is a well-known feature, where a Gaussian-fluctuation *free energy* predicts structure factors that go beyond the so-called "RPA" expressions.[31,87] The correct procedure involves calculating the solution response to an external perturbation, and our preliminary derivations find that corrections to the "RPA" structure factors arise from the perturbation response of $G$ and $\eta$. These results will be reported in future work.

While in this paper we have focused on linear homopolyelectrolyte solutions, our theory is applicable to general polyelectrolyte systems such as polyelectrolyte coacervates, dendrimers, and gels. Promisingly, our theory gives a systematic framework for studying the impact of arbitrary polyelectrolyte architectures on electrostatic correlations, not achievable by many commonly-used theories of polyelectrolyte thermodynamics (i.e. TPT-1 and other theories that use the VO disconnected charges approximation). This feature of our theory is critical for advancing the theoretical design of novel polyelectrolyte materials, for which polymer architecture is a particularly important design parameter.

Finally, our theory retains many of the advantages of the original Gaussian variational theory applied to simple electrolytes,[57] providing a systematic framework for studying inhomogeneities in the dielectric medium and concentration profiles. Study of inhomogeneous systems will be the subject of future work.

**ACKNOWLEDGMENTS**

The authors would like to thank Bilin Zhuang, Rui Wang, Pengfei Zhang and Marco Heinen, for helpful discussions. KS acknowledges support by the NSF-GRFP fellowship.